\documentclass[fleqn,usenatbib]{mnras}

\usepackage{newtxtext,newtxmath}

\usepackage[T1]{fontenc}
\usepackage{ae,aecompl}

\usepackage{graphicx}	
\usepackage{amsmath}	
\usepackage{amssymb}	

\title[Blazar VHE spectral alterations induced by photon-ALP oscillations]{Blazar VHE spectral alterations induced by photon-ALP oscillations}

\author[G. Galanti et al.]{
Giorgio Galanti,$^{1}$\thanks{E-mail: gam.galanti@gmail.com (GG)}
Fabrizio Tavecchio,$^{1}$
Marco Roncadelli$^{2}$
and Carmelo Evoli$^{3}$
\\
$^{1}$INAF, Osservatorio Astronomico di Brera, Via Emilio Bianchi 46, I -- 23807 Merate, Italy\\
$^{2}$INFN, Sezione di Pavia, Via A. Bassi 6, I -- 27100 Pavia, Italy, and INAF\\
$^{3}$GSSI, Viale Francesco Crispi 7, I -- 67100 L'Aquila, Italy, and INFN, LNGS, I -- 67100 Assergi, L'Aquila, Italy
}

\date{Accepted XXX. Received YYY; in original form ZZZ}

\pubyear{2019}

\begin{document}
\label{firstpage}
\pagerange{\pageref{firstpage}--\pageref{lastpage}}
\maketitle

\begin{abstract}
Prompted by the increasing interest of axion-like particles (ALPs) for very-high-energy (VHE) astrophysics, we have considered a full scenario for the propagation of a VHE photon/ALP beam emitted by a BL Lac and reaching us in the light of the most up-to-date astrophysical information and for energies up to above $100 \, \rm TeV$. During its trip, the beam -- generated in a small region of a BL Lac jet -- crosses a variety of magnetic structures in very different astronomical environments: the BL Lac jet, the host elliptical galaxy, the extragalactic space and the Milky Way. We have taken an effort to model all these magnetic fields in the most realistic fashion and using a new model developed by us concerning the extragalactic magnetic field. Assuming an intrinsic spectrum with a power law exponentially truncated at a fixed cut-off energy, we have evaluated the resulting observed spectra of Markarian 501, the extreme BL Lac 1ES 0229+200 and a similar source located at $z = 0.6$ up to above $100 \, \rm TeV$. We obtain interesting results: the model with photon-ALP oscillations possesses features ({\it spectral energy oscillatory behaviour} and {\it photon excess} above $20 \, \rm TeV$) which can be tested by $\gamma$-ray observatories like CTA, HAWC, GAMMA 400, LHAASO, TAIGA-HiSCORE and HERD. In addition, our ALP can be detected in dedicated laboratory experiments like the upgrade of ALPS II at DESY, the planned IAXO and STAX experiments, as well as with other techniques developed by Avignone and collaborators.
\end{abstract}





\section{Introduction}
Everybody knows that the atmosphere is fully opaque to gamma rays. While this is good for us -- otherwise life would be impossible on Earth -- for many years it has been regarded as a stumbling block for gamma-ray observations from the ground. Only about twenty years ago or so has it been realized that the atmospheric opacity can be an opportunity for ground-based gamma-ray observations in the very-high-energy band (VHE, $100 \, {\rm GeV} \lesssim {\cal E} \lesssim 100 \, {\rm TeV})$. Basically, the idea is as follows. When a VHE photon coming from a blazar -- AGN with a jet occasionally pointing towards us -- strikes the atmosphere, it gives rise to a very energetic shower including charged particles and secondary photons. Because the charged particles have a speed slightly higher than the velocity of light in the atmosphere, they produce a flash of violet Cherenkov light about 8 km above the Earth. Such a Cherenkov light can be observed with one or more 10 meter class telescopes, and from the shape of the shower and the specific properties of the Cherenkov light one can infer both the energy and the arrival direction of the primary photon by means of the Imaging Atmospheric Cherenkov Telescopes (IACTs). Nowadays, three of them are operative: H.E.S.S. (High Energy Stereoscopic System)~\citep{hess}, MAGIC (Major Atmospheric Gamma Imaging Cherenkov Telescopes)~\citep{magic} and VERITAS (Very Energetic Radiation Imaging Telescope Array System)~\citep{veritas}, which have detected blazars out to redshift $z \simeq 0.9$ and reach energies at most up to ${\cal O} (10 \,  {\rm TeV})$. But the upcoming CTA (Cherenkov Telescope Array) -- consisting of about fifty IACTs in the south site and about thirty IACTs in the north site -- will be able to probe the whole VHE band with great sensitivity and full sky coverage~\citep{cta}. 

Unfortunately, this kind of observations suffer from another sort of opacity: the {\it extragalactic background light} (EBL). This is the infrared/optical/ultraviolet light emitted by the whole population of galaxies during their cosmic evolution (for a review, see~\citealt{dwek}). As a consequence, a VHE photon of energy ${\cal E}$ can scatter off an EBL photon of energy $\epsilon$, thereby producing an $e^+ e^-$ pair according to the Breit-Wheeler process $\gamma + \gamma \to e^+ + e^-$~\citep{breitwheeler,heitler}. It can be shown that for VHE photons the corresponding cross-section becomes maximal just in the energy range where the EBL dominates~\citep{gs1967}. So, what happens is that the optical depth increases with both $z$ and ${\cal E}$~\citep{stecker1970}, thereby progressively more and more limiting the observations at VHE as either $z$ or ${\cal E}$ (or both) get larger and larger (for an updated quantitative account, see~\citealt{dgr2013}).

A partial way out of this difficulty was first proposed in 2007 in terms of photon-ALP {\it oscillations} in the extragalactic space~\citep{drm2007}. ALPs are axion-like particles -- whose properties will be summarized below -- but the main point is that they totally avoid EBL absorption. In 2008, a complementary possibility has been put forward: photon-to-ALP conversion inside a blazar and ALP-to-photon reconversion in the Galaxy~\citep{shs2008}. 

The aim of this Paper is to contemplate all the magnetic environments at once crossed by the photon/ALP beam -- blazar jet, host galaxy, extragalactic space and Milky Way -- so as to provide the most accurate description as possible according to the present state of the art. A first attempt along the same direction has been done in 2009~\citep{prada1}, using however some simplistic assumptions in the lack of a better knowledge: in this field the progress since 2009 has really been impressive.

As a matter of fact, we identify six points that can and should be improved. 

\begin{itemize}

\item The magnetic field in the blazar jet should be described by an analytic model rather than by a domain-like one as in~\citet{prada1}.

\item Rather than assuming some value for the photon-ALP conversion probability as in~\citet{prada1}, we compute it in terms of realistic values of the relevant parameters.

\item New bounds on the model parameters have been derived.

\item The photon dispersion on the CMB -- whose relevance has been realized in 2015~\citep{raffelt2015} -- plays a crucial role in the description of the extragalactic photon/ALP beam propagation in order to deal with energies up to above $100 \, \rm TeV$, as required by the next generation of gamma-ray detectors (more about this, towards the end of the Paper). 

\item The EBL model has been improved in 2017~\citep{franceschinirodighiero}.

\item The Galactic magnetic field has been modeled in considerable detail by~\citet{jansonfarrar1,jansonfarrar2}.

\end{itemize}

Thus, we will evaluate the total VHE photon survival probability within the full scenario, namely from the origin in the BL Lacs jet, during the propagation inside the jet, within the host galaxy and the extragalactic space, and finally inside the Milky Way to us by taking the above six points into account. In addition, assuming a realistic emitted spectrum for three BL Lacs -- Markarian 501, 1ES 0229+200 and a similar source located at $z=0.6$ -- we derive the observed spectrum up to above $100 \, \rm TeV$.

\section{General features of axion-like particles (ALPs)}

As we said, a key-role is played in our considerations by axion-like particles (ALPs). They are attracting growing interest, especially because they are a natural prediction of superstring theories (for a review, see~\citealt{alp1,alp2}). Let us very cursorily recall their most relevant properties (more about them in~\citealt{grSM,grExt}).

They are spin-zero, neutral and extremely light pseudo-scalar bosons. As far as our purposes are concerned, they are described by the Lagrangian
\begin{equation}
\label{t1}
{\cal L}_{\rm ALP} = \frac{1}{2} \, \partial^{\mu} a \, \partial_{\mu} a - \, \frac{1}{2} \, m_a^2 \, a^2 + g_{a \gamma \gamma} \, a \, {\bf E} \cdot {\bf B}~,
\end{equation}
where ${\bf E}$ and ${\bf B}$ denote the electric and magnetic components of the electromagnetic tensor $F^{\mu \nu}$ and $a$ stands for the ALP field. Moreover, the two-photon coupling of $a$ -- namely $g_{a \gamma \gamma}$ -- is totally unrelated to the ALP mass $m_a$.

We shall henceforth consider a photon/ALP beam of VHE ${\cal E}$ which propagates from a BL Lac towards us along the $y$ direction, in the presence of an {\it external} magnetic field ${\bf B}$ which depends on the considered environment. So, the mass matrix of the $\gamma - a$ system is off-diagonal (the electric field ${\bf E}$ in Eq. (\ref{t1}) pertains to a propagating  VHE photon). As a consequence, the propagation eigenstates differ from the interaction eigenstates: hence $\gamma \leftrightarrow a$ {\it oscillations} take place, since ALPs have no chance to decay since for values of $m_a$ to be considered below the time for the process $a \to \gamma + \gamma$ is much longer than the age of the Universe~\citep{sikivie,anselm,rs1988}. Because ${\bf E}$ is orthogonal to the photon momentum ${\bf k}$, only the component ${\bf B}_T$ of ${\bf B}$ -- transverse to ${\bf k}$ --  couples to $a$.
     
When ${\bf B}$ is strong -- like in the case of the jet of BL Lacs -- also the one-loop QED vacuum polarization must be taken into account, which is described by the effective Lagrangian~\citep{rs1988,HEW1,HEW2,HEW3}
\begin{equation}
\label{HEW}
{\cal L}_{\rm HEW} = \frac{2 \alpha^2}{45 m_e^4} \, \left[ \left({\bf E}^2- {\bf B}^2 \right)^2+7 \left( {\bf E}\cdot{\bf B} \right)^2 \right]~,
\end{equation}
where $\alpha$ is the fine-structure constant and $m_e$ is the electron mass. Note that ${\cal L}_{\rm HEW}$ holds true for ${\cal E} \gg m_e$~\citep{raffelt2015}. In addition, for weak magnetic fields -- namely the ones permeating extragalactic space ${\cal O}(1 \, \rm nG)$ (more about this later) -- and for ${\cal E} \gtrsim {\cal O}(10 \, \rm TeV)$ the photon dispersion on the cosmic microwave background (CMB) plays a leading role~\citep{raffelt2015}, and we will include it (the rationalized system with natural units will systematically be employed). 

From Eqs. (\ref{t1}) and (\ref{HEW}) it is possible to derive the propagation equation for the photon/ALP beam. Actually, since (as we shall see) we will be dealing with the situation  $m_a \ll {\cal E}$, such an equation reduces to a Schr\"odinger-like equation with time replaced by $y$~\citep{rs1988}: see e.g.~\citet{grSM,dgrx} to which we refer the reader for more details. Thus, the photon/ALP beam is formally described as a three-level ($A_x(y)$, $A_z(y)$ and $a(y)$) non-relativistic unstable quantum system, where $A_x(y)$ and $A_z(y)$ denote the photon amplitudes with polarization along the $x$ and $z$ axis, respectively, while $a(y)$ is the amplitude associated with the ALP. Accordingly, the most important quantity which quantifies the beam propagation is the {\it transfer matrix} of the Schr\"odinger-like equation, which will be referred to as ${\cal U}$: it is the solution ${\cal U} ({\cal E}; y, y_0)$ such that 
${\cal U} ({\cal E}; y_0, y_0) = 1$.

\subsection{Parameter space}
Before plunging into our analysis, it is obviously compelling to know the allowed ranges of the two photon coupling $g_{a \gamma \gamma}$ and the ALP mass $m_a$.

A thorough discussion of the various bounds delimiting the allowed parameter space is contained in~\citet{grExt}, and here we merely report the relevant ones. 

\begin{itemize} 

\item $g_{a \gamma \gamma} < 0.66 \cdot 10^{- 10} \, {\rm GeV}^{- 1}$ for $m_a < 0.02 \, {\rm eV}$ at the $2 \sigma$ level from the lack of detection of ALPs coming from the Sun~\citep{cast} and from stellar evolution of certain stars in globular clusters~\citep{straniero}. 

\item $g_{a \gamma \gamma} \lesssim 5 \cdot 10^{- 12} \, {\rm GeV}^{- 1}$ for $5 \cdot 10^{- 10} \, {\rm eV} \lesssim m_a  \lesssim 5 \cdot 10^{- 9} \, {\rm eV}$ at the $2 \sigma$ level from observations of the Perseus cluster~\citep{fermi2016}.

\item $g_{a \gamma \gamma} \lesssim 5.3 \cdot 10^{- 12} \, {\rm GeV}^{- 1}$ for $m_a \lesssim 4.4 \cdot 10^{- 10} \, {\rm eV}$ from the lack of detection of gamma-rays from supernova SN1987A~\citep{payez2015}.

\end{itemize} 

We take into account the first bound. The second bound can be met provided that we assume $m_a = {\cal O}(10^{- 10} \, {\rm eV})$\footnote{Our results are not strongly affected by the value of $m_a$ so that we prefer to give for $m_a$ an order of magnitude in order to maintain generality.}. Because of the strong criticism discussed in great detail elsewhere~\citep{prada1,grExt}, we do not trust the third bound. In any case, we have tried to restrict our model to the third bound\footnote{In this fashion it is possible to take $m_a \sim 4.5 \cdot 10^{-10} \, \rm eV$ avoiding limitations on $g_{a\gamma\gamma}$.} and the effects and features discussed below qualitatively remain. 

Thus, in order to be specific we choose as benchmark values  $g_{a \gamma \gamma} = 10^{- 11} \, {\rm GeV}^{- 1}$ and $m_a = {\cal O}(10^{- 10} \, {\rm eV})$.

\section{Photon-ALP beam propagation}

In this section we study the propagation of the photon/ALP beam starting at the BL Lac jet base up to its arrival at the Earth. 

\subsection{Photon/ALP beam propagation in the jet}

We denote by ${\cal R}_{\rm VHE}$ the region where the VHE photons originate inside the BL Lac jet, letting $y_{\rm VHE}$ be its distance from the central black hole (BH). So, our first step is to evaluate the transfer matrix over the jet region ${\cal R}_{\rm jet}$ between $y_{\rm VHE}$ and the end of the jet $y_{\rm jet}$, which we denote as ${\cal U}_{{\cal R}_{\rm jet}} ({\cal E}; y_{\rm jet}, y_{\rm VHE})$.

Here, we closely follow our results concerning the same problem as derived in a previous Letter~\citep{trg}. We start by recalling that the region ${\cal R}_{\rm VHE}$ is rather far from the central BH, and the jet axis is supposed to coincide with the direction $y$. In order to evaluate the photon/ALP beam propagation inside the jet we must know three quantities: 1) the distance $y_{\rm VHE}$ from the central BH, 2) the transverse magnetic field profile $B_{T, {\cal R}_{\rm jet}}  ( y )$ from $y_{\rm VHE}$ to $y_{\rm jet}$, 3) the electron density profile $n_{e, {\cal R}_{\rm jet}} ( y )$ from $y_{\rm VHE}$ to $y_{\rm jet}$. Realistic values for these quantities can be derived from Synchrotron Self Compton (SSC) diagnostics as applied to the {\it spectral energy distribution} (SED) of BL Lacs~\citep{tavecchio2010}. Inside ${\cal R}_{\rm VHE}$ we get $B_{T, {{\cal R}_{\rm VHE}}} = (0.1 - 1) \, {\rm G}$ and in order to be definite we choose $B_{T, {{\cal R}_{\rm VHE}}} = 0.5 \, {\rm G}$. Moreover, we find $n_{e, {{\cal R}_{\rm VHE}}} \simeq 5 \cdot 10^{4} \, {\rm cm}^{-3}$, leading in turn to a plasma frequency of $\omega_{\rm pl} \simeq 8.25 \cdot 10^{- 9} \, {\rm eV}$. Although there is no direct way to infer a precise value of $y_{\rm VHE}$, we can estimate it from the size of ${\cal R}_{\rm VHE}$ -- which is assumed to be a measure of the jet cross-section --  thus finding $y_{{\rm VHE}} = (10^{16} - 10^{17}) \, {\rm cm}$. For definiteness, we shall take $y_{\rm VHE} \simeq 3 \cdot 10^{16} \, {\rm cm}$. Once produced, VHE photons propagate unimpeded out to $y_{\rm jet} \simeq 1 \, \rm kpc$ where they leave the jet, entering the host galaxy. More specifically, within ${\cal R}_{\rm jet}$ what is relevant is the toroidal part of the magnetic field which is transverse to the jet axis~\citep{bbr1984,ghisellini2009,pudritz2011}. Its profile reads
\begin{equation}
\label{Bjet}
B_{T, {\cal R}_{\rm jet}} ( y ) = B_{T, {\cal R}_{\rm VHE}} \left(\frac{y_{{\rm VHE}}}{y}\right)~.
\end{equation}
Concerning the electron density profile, due to the conical shape of the jet our expectation is
\begin{equation}
\label{njet}
n_{e, {{\cal R}_{\rm jet}}} ( y ) = n_{e, {{\cal R}_{\rm VHE}}}  
\left(\frac{y_{{\rm VHE}}}{y}\right)^2~.
\end{equation}
By knowing the above quantities, it is possible to calculate the entire propagation process of the photon/ALP beam within the jet, namely ${\cal U}_{{\cal R}_{\rm jet}}({\cal E}; y_{\rm jet}, y_{\rm VHE})$. 

Some remarks are in order. In the first place, we have provided a detailed modeling of the magnetic field in a BL Lac jet, which greatly differs from the domain-like one employed in the original full scenario for PKS 2155-304~\citep{prada1}. Moreover, we stress that in ${\cal R}_{\rm jet}$ we consider the photon/ALP beam in a frame co-moving with the jet, so that we must apply the transformation ${\cal E} \to \gamma {\cal E}$ to the beam in order to go to a fixed frame -- as we will do in the next regions -- with $\gamma$ being the Lorentz factor. We take $\gamma=15$.

\subsection{Photon/ALP beam propagation in the host galaxy}

All the three considered blazars are hosted by an elliptical galaxy, which we denote by ${\cal R}_{\rm host}$. 
According to the common wisdom, the magnetic field inside them is turbulent and described by a domain-like model with average strength ${B}  \simeq 5 \, \mu{\rm G}$ and coherence length $L_{\rm dom} \simeq 150 \, {\rm pc}$~\citep{moss1996}. As we have shown elsewhere~\citep{trgb2012}, its effect on the photon/ALP beam is totally negligible -- since the $\gamma \leftrightarrow a$ oscillation length is much larger than $L_{\rm dom}$ -- and so we ignore it. Therefore, denoting by $y_{\rm in, host} \equiv y_{\rm jet}$ and by $y_{\rm out, host}$ the points on the $y$ axis where the beam enters and exits from the host galaxy, respectively, we have ${\cal U}_{{\cal R}_{\rm host}} ({\cal E}; y_{\rm out, host}, y_{\rm in, host}) = 1$.

\subsection{Photon/ALP beam propagation in the extragalactic space}

We let ${\cal R}_{\rm ext}$ be the region where the photon/ALP beam propagates in the extragalactic space, i.e. from $y_{\rm out, host}$ up to the border of the Milky Way $y_{\rm MW}$. Clearly, the beam behaviour in ${\cal R}_{\rm ext}$ is affected by the morphology and strength of the extragalactic magnetic field ${\bf B}_{\rm ext}$. Unfortunately, almost nothing is known about it, and several configurations for ${\bf B}_{\rm ext}$ have been proposed~\citep{kronberg1994,grassorubinstein,wanglai,jap}. Current limits restrict $B_{\rm ext}$ to the range $10^{- 7} \, {\rm nG} \leq {B}_{\rm ext} \leq 1.7 \, {\rm nG}$ on the scale of ${\cal O} (1 \, {\rm Mpc})$~\citep{neronov2010,durrerneronov,pshirkov2016}. 

According to the current wisdom, ${\bf B}_{\rm ext}$ is modeled as a domain-like network, in which ${\bf B}_{\rm ext}$ is assumed to be homogeneous over a whole domain of size $L_{\rm dom}$ equal to its coherence length, with ${\bf B}_{\rm ext}$ changing randomly and {\it discontinuously} its direction from one domain to the next but keeping approximately the same strength~\citep{kronberg1994,grassorubinstein}. We remark that this scenario relies upon outflows from primeval galaxies, further amplified by turbulence~\citep {reessetti,hoyle,kronberg1999,furlanettoloeb}. Common benchmark values are 
$B_{\rm ext} = {\cal O}(1 \, \rm nG)$ on a coherence length ${\cal O}(1 \, \rm Mpc)$, thereby implying that the size of the magnetic domains is $L_{\rm dom} = {\cal O}(1 \, \rm Mpc)$. A careful analysis of the motivation behind this model has been provided in~\citet{grSM}. In order to be definite, we choose $B_{\rm ext} = 1 \, {\rm nG}$.

What about the above {\it unphysical} jumps of $B_{\rm ext}$ across the domain edges? As discussed in great detail in~\citet{grSM}, what ultimately matters is the $\gamma \leftrightarrow a$ oscillation length $l_{\rm osc}$. Somewhat schematically, the situation can be summarized as follows. 

\begin{itemize} 

\item So long as $l_{\rm osc} \gg \, L_{\rm dom}$, the standard model for ${\bf B}_{\rm ext}$ as outlined above is perfectly viable. This is the typical situation encountered so far, since presently operating IACTs reach at most energies up to ${\cal E} = {\cal O} (10 \, {\rm TeV})$, for which we indeed have $l_{\rm osc} \gg \, L_{\rm dom}$. We stress that 
one of its advantages is that the beam propagation equation inside a single domain is very easy to solve. 

\item But in this Paper we are interested in energies up to above $100 \, {\rm TeV}$, and this fact brings about a big difference. As mentioned previously, for ${\cal E} \gtrsim {\cal O}(10 \, \rm TeV)$ the photon dispersion on the CMB~\citep{raffelt2015} becomes dominant, and it must be include into our scenario. Moreover, it has been shown that -- because of such an effect -- $l_{\rm osc}$ decreases as ${\cal E}$ increases, and in particular it is found that  $l_{\rm osc} \lesssim {\cal O} (1 \, {\rm Mpc})$ for ${\cal E} \gtrsim {\cal O} 
(40 \, {\rm TeV})$~\citep{grExt,raffeltvogel}.

\end{itemize} 

As a consequence, a more accurate domain-like model for ${\bf B}_{\rm ext}$ has to be developed, in which the variation of ${\bf B}_{\rm ext}$ is {\it smooth} when passing from a domain to the next~\citep{grSM,raffeltvogel}. Although there are many ways to implement such an idea, a {\it linear smoothing} of the domain edges seems the simplest possibility, which has been worked out in full detail in~\citet{grSM}. We denote by $\sigma$ the smoothing parameter which measures the path that the photon/ALP beam spends in the smoothed region of a single domain: e.g. $\sigma=0.2$ means that the beam propagates in the constant angle region for $80 \%$ of its path and in the smoothly varying angle region for $20 \%$. In order to be definite, we shall take $\sigma = 0.2$. We also let the length of $L_{\rm dom}$ vary according to a power law distribution function $\propto L_{\rm dom}^{- 1.2}$ inside the range $0.2 \, {\rm Mpc} - 10 \, {\rm Mpc}$, so that $\langle L_{\rm dom} \rangle = 2 \, {\rm Mpc}$ -- which is in agreement both with the considered physical scenario and with the present bounds~\citep{durrerneronov}. 

Finally, we emphasize that -- because of the random direction of ${\bf B}_{\rm ext}$ in every domain -- the photon/ALP beam propagation becomes a stochastic process, and so what we actually observe is only a {\it single realization} of that process.

Coming back to photon/ALP beam propagation in ${\cal R}_{\rm ext}$, it is discussed within the theoretical framework developed in~\citet{grSM,grExt}, where the most recent data about the extragalactic background light (EBL) are employed~\citep{franceschinirodighiero}. Incidentally, a somewhat different approach~\citep{kk2017} has been based on the EBL model obtained by the CIBER experiment~\citep{ciber}. Accordingly, we denote by ${\cal U}_{{\cal R}_{\rm ext}}({\cal E}; y_{\rm MW}, y_{\rm out, host})$ the corresponding transfer matrix of the photon/ALP beam.

Owing to $\gamma \leftrightarrow a$ oscillations, photons acquire a split personality: when they propagate like true photons they suffer EBL absorption, but when they propagate as ALPs absorption is totally absent. As a result, the effective optical depth $\tau_{\rm ALP} ({\cal E}, z)$ is smaller than in conventional physics. The gist of the argument is that the photon survival probability is now
\begin{equation}
\label{njet}
P^{\rm ALP}_{\gamma \to \gamma} ({\cal E}, z) = {\rm exp} \, \bigl[- \, \tau_{\rm ALP} ({\cal E}, z) \bigr]~,
\end{equation}
and even a small decrease of $\tau_{\rm ALP} ({\cal E}, z)$ produces a large increase of $P^{\rm ALP}_{\gamma \to \gamma} ({\cal E}, z)$ as compared to conventional physics.

\subsection{Photon/ALP beam propagation in the Milky Way}
We denote by ${\cal R}_{\rm MW}$ the region where the photon/ALP beam propagates inside the Milky Way, i.e. from $y_{\rm MW}$ up to the Earth, whose position is denoted by $y_\oplus$.

We compute ${\cal U}_{{\cal R}_{\rm MW}} ({\cal E}; y_\oplus, y_{\rm MW})$ by closely following the strategy described in~\citet{hmmmmr2012}. Specifically, in order to take into account the structured behaviour of the Galactic magnetic field ${\bf B}_{\rm MW}$ we adopt the recent Jansson and Farrar model~\citep{jansonfarrar1,jansonfarrar2}, which includes a disk and a halo component, both parallel to the Galactic plane, and poloidal `X-shaped' component at the galactic center. Its latest updated version is described in~\citet{uf2017}, where newer polarized synchrotron data and use of different models of the cosmic ray and thermal electron distribution are exploited.

The other model of the Galactic magnetic field existing in the literature is the one by~\citet{pshirkovMF2011}: however, this model is based mainly on data along the Galactic plane so that the Galactic halo component of ${\bf B}_{\rm MW}$ is not determined with accuracy. For this reason we prefer to use the Jansson and Farrar model. In any case, we have tested the robustness of our findings by employing also this model and even if with some little modifications our results are qualitatively unchanged. 

While the Jansson and Farrar model allows also for a random and a striated component of the field, it turns out that only the regular component is relevant in the present context, since the $\gamma \leftrightarrow a$ oscillation length is much larger than the coherence length of the turbulent field. 

Inside the Milky Way disk the electron number density is $n_e \simeq 1.1 \cdot 10^{-2} \, {\rm cm}^{-3}$, resulting in a plasma frequency $\omega_{\rm pl} \simeq 3.9 \cdot 10^{-12} \, {\rm eV}$: this emerges from a new model for the distribution of the free electrons in the Galaxy~\citep{ymw2017}. The Galaxy is modeled by an extended thick disk accounting for the so-called warm interstellar medium, a thin disk standing for the Galactic molecular ring, spiral arms (inferred from a new fit to Galactic HII regions), a Galactic Center disk and seven local features counting the Gum Nebula, the Galactic Loop I and the Local Bubble. The model includes an offset of the Sun from the Galactic plane and a warp of the outer Galactic disk. The Galactic model parameters are obtained by fitting to 189 pulsars with independently determined distances and DMs. 

Thanks to this procedure, we can compute ${\cal U}_{{\cal R}_{\rm MW}}({\cal E};y_\oplus,y_{\rm MW})$ for an arbitrary direction of the line of sight to a given blazar.

\subsection{Overall photon survival probability}

Once all transfer matrices in each region are known, the total transfer matrix $\cal U$ describing the propagation of the photon/ALP beam from the VHE photon production region in the BL Lac jet up to the Earth reads 
\begin{eqnarray} 
\label{Utot}
&\displaystyle {\cal U}({\cal E}; y_\oplus, y_{\rm VHE}) = {\cal U}_{{\cal R}_{\rm MW}} ({\cal E}; y_\oplus, y_{\rm MW}) \times \\ \nonumber
&\displaystyle {\cal U}_{{\cal R}_{\rm ext}}({\cal E}; y_{\rm MW}, y_{\rm out, host}) \, {\cal U}_{{\cal R}_{\rm host}} ({\cal E}; y_{\rm out, host}, y_{\rm in, host}) \times \\ \nonumber
&\displaystyle {\cal U}_{{\cal R}_{\rm jet}}({\cal E}; y_{\rm in, host}, y_{\rm VHE})~, \nonumber
\end{eqnarray}
where of course we have $y_{\rm in, host} \equiv y_{\rm jet}$. Since photon polarization cannot be measured in the VHE gamma-ray band, we have to treat the beam as unpolarized. Therefore, we must use the generalized polarization density matrix $\rho(y)=(A_x(y),A_z(y),a(y))^T \otimes (A_x(y),A_z(y),a(y))$. As a consequence, the overall photon survival probability takes the form
\begin{eqnarray} 
\label{prob}
&\displaystyle P^{\rm ALP}_{\gamma \to \gamma} \bigl({\cal E}; y_\oplus, \rho_x, \rho_z; y_{\rm VHE}, \rho_{\rm unp} \bigr) = \\
&\displaystyle \sum_{i = x,z} {\rm Tr} \left[\rho_i \, {\cal U} \bigl({\cal E}; y_\oplus, y_{\rm VHE} \bigr) \, \rho_{\rm unp} \, {\cal U}^{\dagger} \bigl({\cal E}; y_\oplus, y_{\rm VHE} \bigr) \right]~,\nonumber
\end{eqnarray}
where $\rho_x \equiv {\rm diag} \, (1,0,0)$, $\rho_z \equiv {\rm diag} \, (0,1,0)$ and 
$\rho_{\rm unp} \equiv {\rm diag} \, (0.5,0.5,0)$. Below -- merely for notational convenience -- we shall replace $P^{\rm ALP}_{\gamma \to \gamma} \bigl({\cal E}; y_\oplus, \rho_x, \rho_z; y_{\rm VHE}, \rho_{\rm unp} \bigr)$ simply by $P^{\rm ALP}_{\gamma \to \gamma} \bigl({\cal E}, z \bigr)$. 

\begin{figure}    
\begin{center}
\includegraphics[width=.49\textwidth]{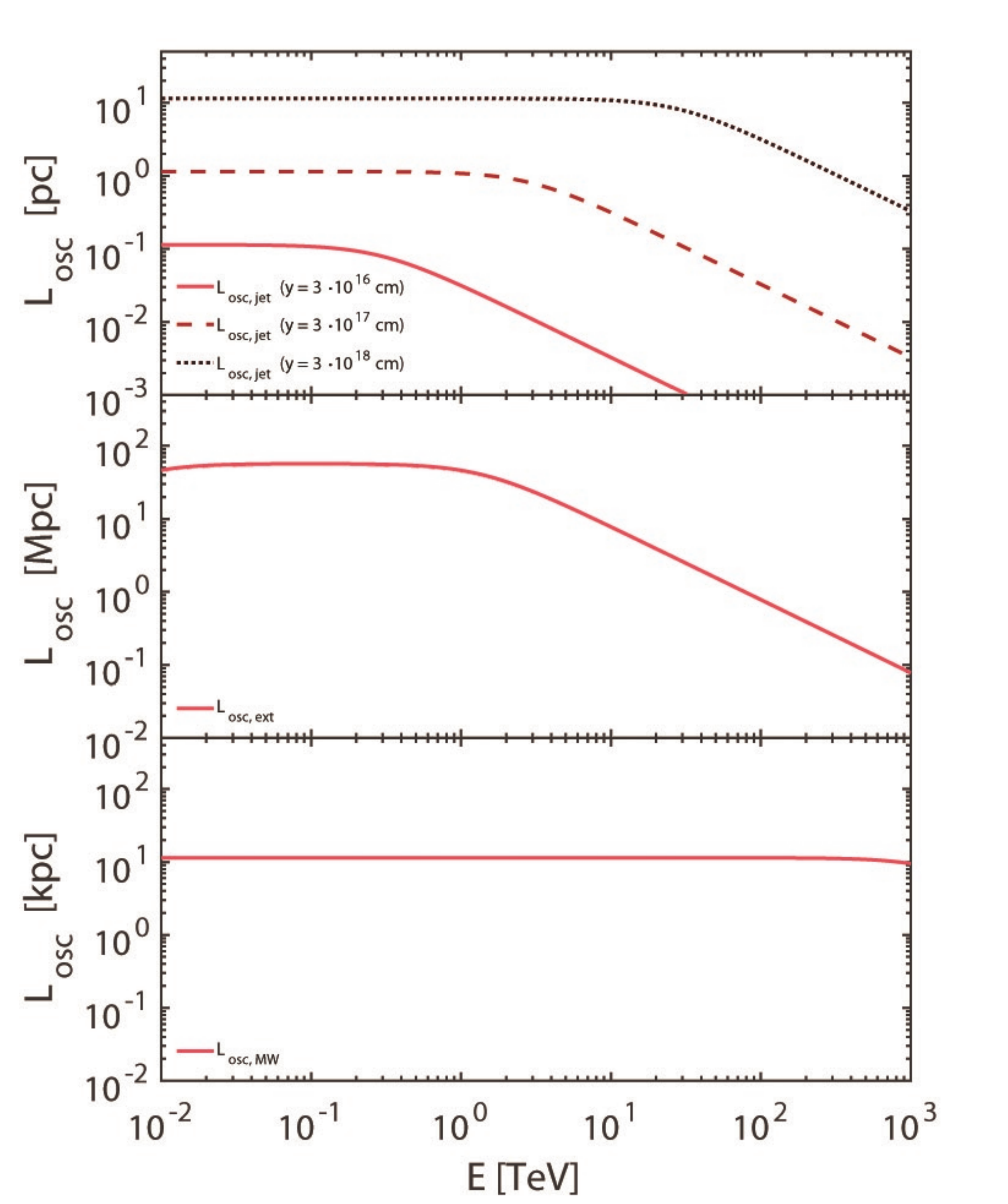}
\end{center}
\caption{\label{Losc} 
Behaviour of the oscillation length $l_{\rm osc}$ versus the observed energy $\cal E$ in the various regions crossed by the photon/ALP beam. The upper panel refers to the propagation inside the jet: in this case $l_{\rm osc}$ strongly depends on the value of $B_{T,{\cal R}_{\rm jet}}(y)$ at different distances from the emission region. We plot $l_{\rm osc}$ at (i) the emission distance $y=y_{\rm VHE}=3 \cdot 10^{16} \, \rm cm$ (solid line), (ii) $y=10\,y_{\rm VHE}=3 \cdot 10^{17} \, \rm cm$ (dashed line) and (iii) $y=100\,y_{\rm VHE}=3 \cdot 10^{18} \, \rm cm$ (dotted line). In the central panel we draw the behaviour of $l_{\rm osc}$ versus $\cal E$ in the extragalactic space while in the lower panel the behaviour of $l_{\rm osc}$ versus $\cal E$ in the Milky Way.}
\end{figure}

In order to give the reader a feeling of what happens in the various regions crossed by the photon/ALP beam, in Figure~\ref{Losc} we plot how the oscillation length $l_{\rm osc}$ varies with the energy $\cal E$ in the jet, in the extragalactic space and in the Milky Way. As the upper panel of Figure~\ref{Losc} shows the behaviour of $l_{\rm osc}$ versus $\cal E$ is strongly affected by the value of $B_{T,{\cal R}_{\rm jet}}(y)$: we observe that as expected (see Eq. (21) of~\citealt{grSM}) as $B_{T,{\cal R}_{\rm jet}}(y)$ decreases (when the distance from the emission region increases) the maximal value of $l_{\rm osc}$ grows and the energy where the QED vacuum polarization effect is important grows as well. Instead, in the central panel of Figure~\ref{Losc} we observe that in the extragalactic space $l_{\rm osc}$ starts to decrease because of the effect of the photon dispersion on the CMB which becomes more and more important as $\cal E$ grows (for more details see~\citealt{grExt}). In the lower panel of Figure~\ref{Losc} we see that in the Milky Way $l_{\rm osc}$ is almost constant as respect to $\cal E$ since the QED vacuum polarization effect and that of the photon dispersion on the CMB are subdominant as respect to the photon-ALP mixing one in almost all the considered energy range.

\section{Blazar spectra}

Starting from the intrinsic spectra, we are now in position to use the overall photon survival probability in order to derive the observed spectra of some blazars -- Markarian 501, 1ES 0229+200 and a similar source located at $z=0.6$ -- and from them to infer the corresponding SED $\nu F_{\nu}$ in the presence of $\gamma \leftrightarrow a$ oscillations all the way from inside the blazar to us. We can thus compare our findings with the results from conventional physics.  

As a preliminary step, we define the (intrinsic or observed) blazar photon spectrum as 
\begin{equation}
\label{mr03112018a}
{\cal F} ({\cal E}) \equiv \frac{dN}{dt dA d {\cal E}}~,
\end{equation}
where $N$ is the VHE photon number and $dA$ is an infinitesimal area. 

For the three considered blazars, we model their intrinsic spectrum with a power law exponentially truncated at a fixed cut-off energy ${\cal E}_{\rm cut}$ as
\begin{equation}
\label{int}
{\cal F}_{\rm int} ({\cal E}) = {\cal F}_0 \, \left( \frac{\cal E}{{\cal E}_0} \right)^{- k} e^{-{\cal E}/{{\cal E}_{\rm cut}}}~,
\end{equation}
where ${\cal F}_0$ is a normalization constant accounting for the blazar luminosity, ${\cal E}_0$ is a reference energy and $k$ is a spectral index. By means of $P^{\rm ALP}_{\gamma \to \gamma}$ in Eq. (\ref{prob}) the observed blazar spectrum turns out to be
\begin{equation}
\label{obs}
{\cal F}_{\rm obs} ({\cal E}) = P^{\rm ALP}_{\gamma \to \gamma} \bigl({\cal E} \bigr) \, {\cal F}_{\rm int} ({\cal E})~.
\end{equation}
In Eq. (\ref{prob}) we use our benchmark values of the free parameters, namely $g_{a \gamma \gamma}=10^{-11} \, {\rm GeV^{-1}}$, $B_{T, {{\cal R}_{\rm VHE}}} = 0.5 \, {\rm G}$, $B_{\rm ext}= 1 \, {\rm nG}$ and $m_a = {\cal O}(10^{ - 10} \, {\rm eV})$.

Before proceeding further, we recall that the SED is related to ${\cal F}_{\rm obs} ({\cal E})$ by
\begin{equation}
\label{mr02112018a}
\nu F_{\nu} = {\cal E}^2 \, {\cal F}_{\rm obs} ({\cal E})~.
\end{equation}

The observable physical quantity is the blazar spectrum pertaining to a single random realization of the photon/ALP propagation process. Nevertheless, it is enlightening to evaluate several realizations at once and to compute some of their statistical properties -- the median and the area containing the $68 \%$, $90 \%$ and $99 \%$ of the total number of realizations -- in order to check the stability of the result against the distribution of the 
${\bf B}_{\rm ext}$ orientation angles and of $L_{\rm dom}$, which are indeed the independent random variables.

\begin{figure*}   
\begin{center}
\includegraphics[width=.9\textwidth]{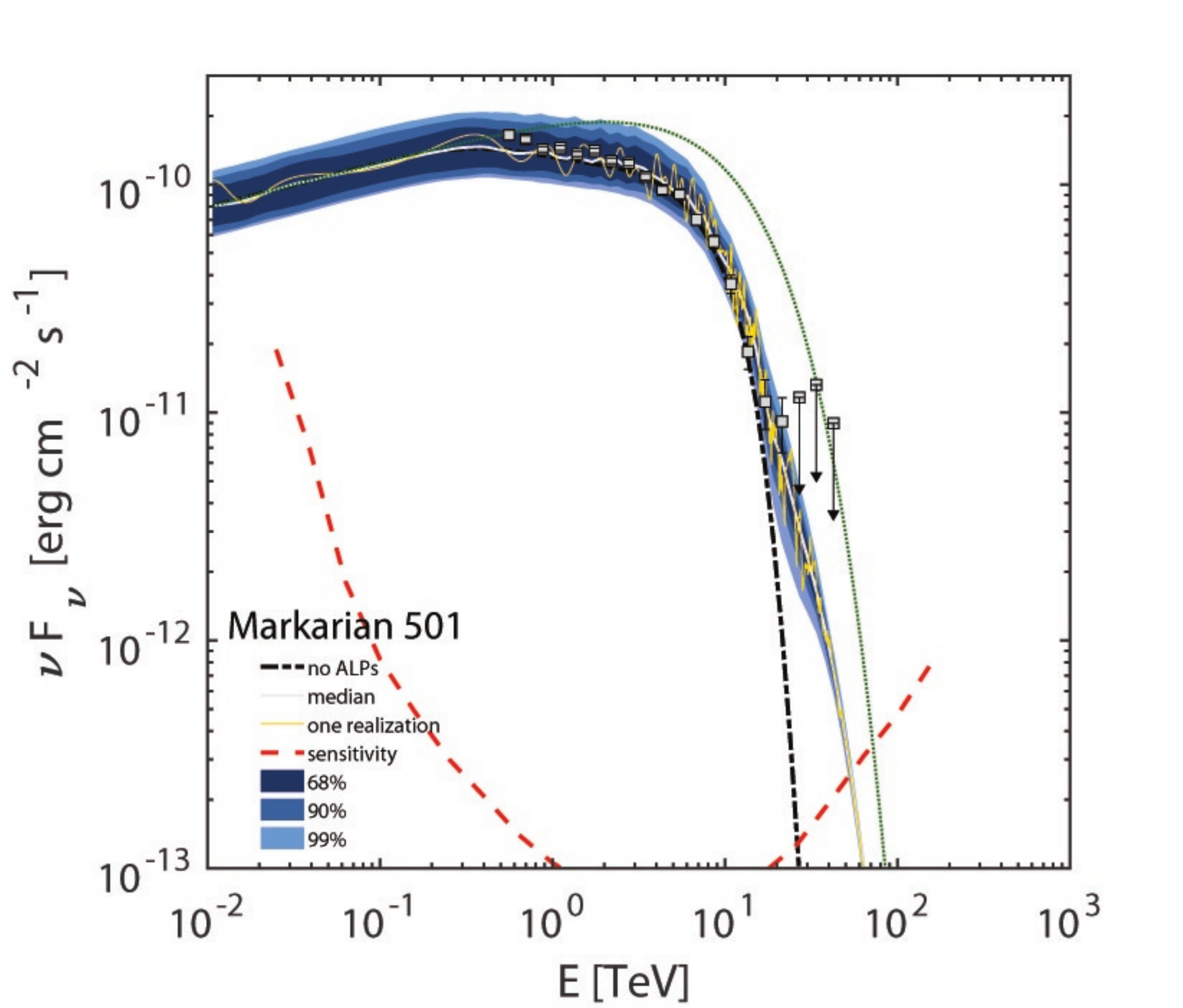}
\end{center}
\caption{\label{Mrk501} 
Behaviour of the observed SED of Markarian 501 versus the observed energy $\cal E$. The dotted-dashed black line corresponds to conventional physics, the solid light-gray line to the median of all the realizations of the photon/ALP propagation process and the solid yellow line to a single realization with a random distribution of the domain lengths and of the orientation angles of the extragalactic magnetic field. The dotted green line is the intrinsic SED and the dashed red line represents the CTA sensitivity for the South site and 50 h of observation. The filled area is the envelope of the results on the percentile of all the possible realizations of the propagation process at $68 \%$ (dark blue), $90 \%$ (blue) and $99 \%$ (light blue), respectively. The light gray squares are the spectrum detected by HEGRA~\citep{hegra}.}
\end{figure*}

\begin{figure*} 
\begin{center}
\includegraphics[width=.9\textwidth]{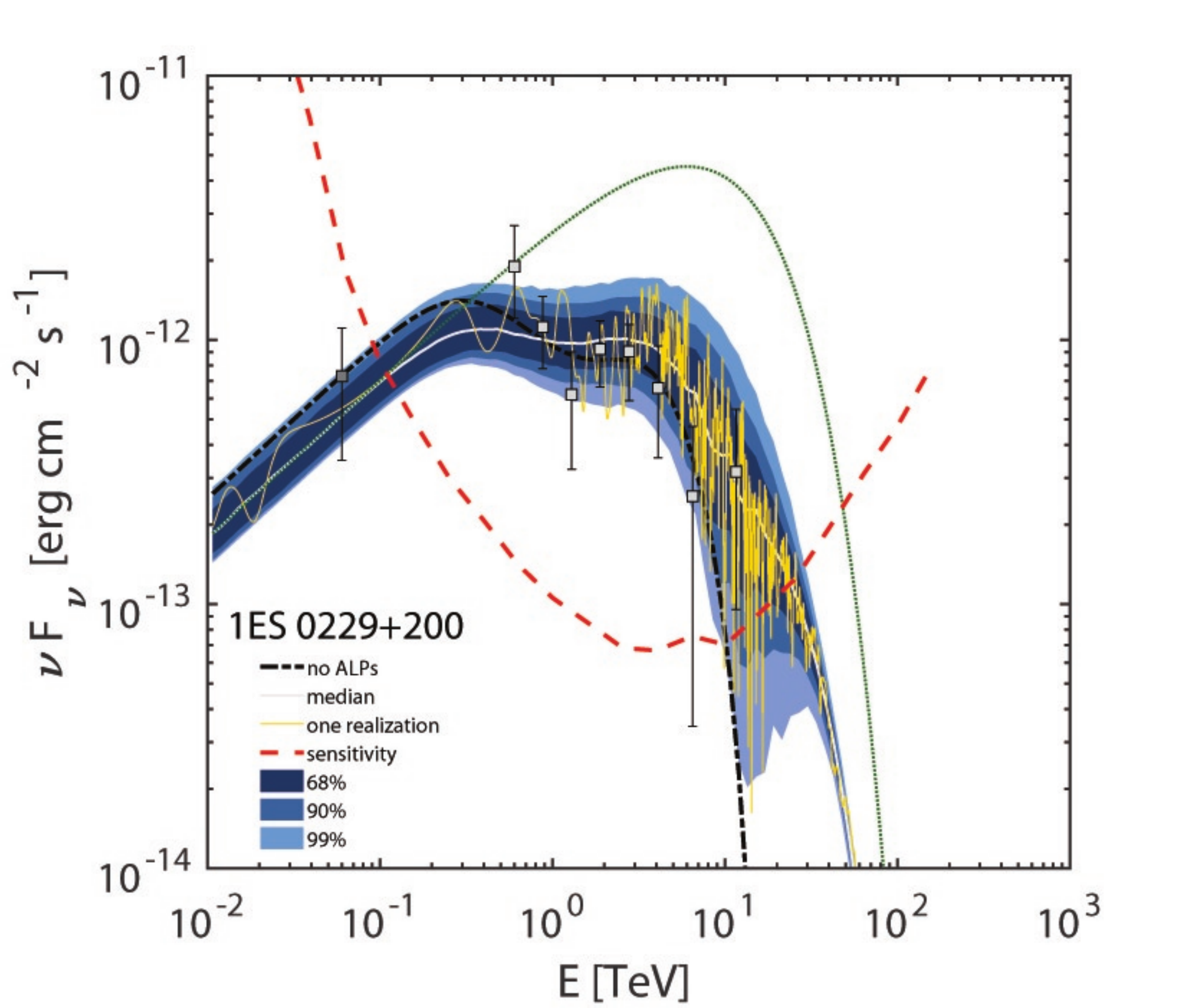}
\end{center}
\caption{\label{0229} 
Same as Figure~\ref{Mrk501} but for 1ES 0229+200. The dark gray squares are the spectrum detected by Fermi/LAT~\citep{1esfermi} while the light gray squares are the spectrum observed by HESS~\citep{1eshess}.}
\end{figure*}

\begin{figure*}    
\begin{center}
\includegraphics[width=.9\textwidth]{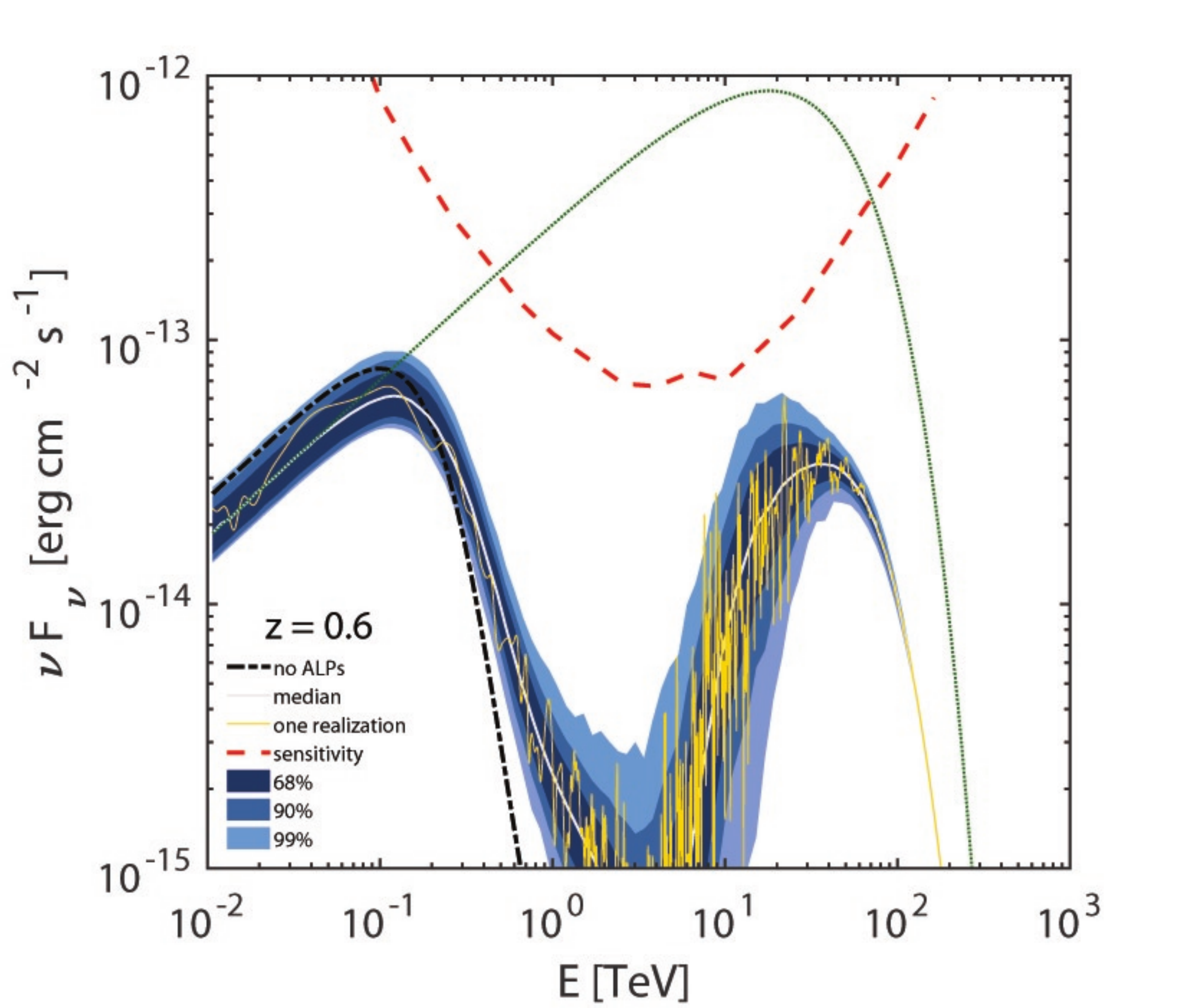}
\end{center}
\caption{\label{z06L} 
Same as Figure~\ref{Mrk501} but for a BL Lac at $z=0.6$ in the case of observation of the BL Lac along the direction of the galactic pole.}
\end{figure*}

\begin{figure*}
\begin{center}
\includegraphics[width=.9\textwidth]{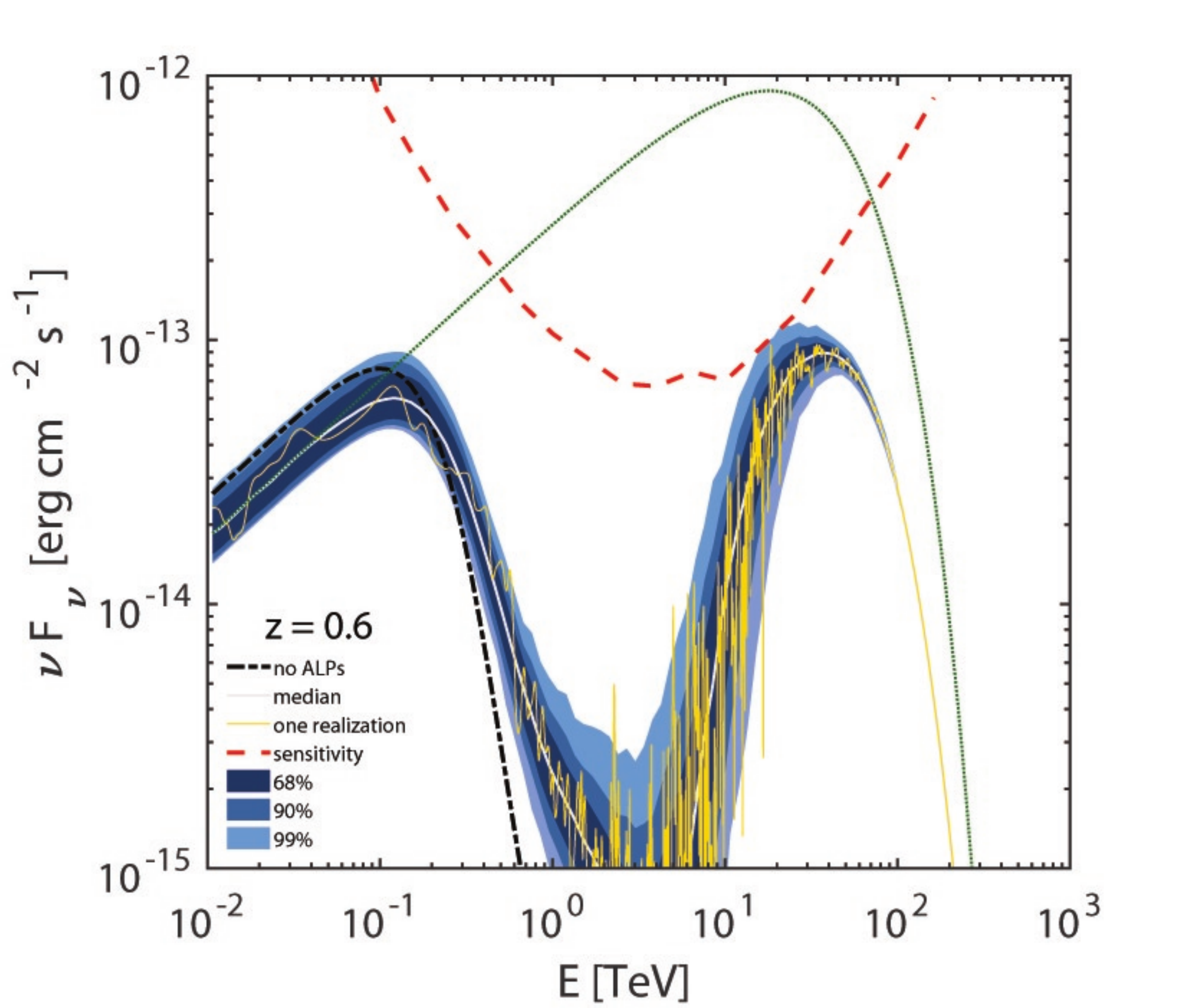}
\end{center}
\caption{\label{z06H} 
Same as Figure~\ref{Mrk501} but for a BL Lac at $z=0.6$ in the case of observation of the BL Lac along the direction of the galactic plane.}
\end{figure*}

\begin{itemize}

\item {\it Markarian 501} -- Markarian 501 is a high-frequency peaked blazar (HBL) observed in the sky at $\rm RA: 16\,h\,53\,m\,52.2\,s$ and $\rm DEC: +39\,d\,45\,m\,37\,s$ at a redshift $z=0.034$. We use the observational data points from HEGRA~\citep{hegra} in a condition where Markarian 501 was observed in a high emission state, thereby allowing us to have a very good description of its spectrum up to $\sim$ 30 TeV. This fact is important for testing our model, since at such high energies it starts to give different predictions with respect to conventional physics. In Figure~\ref{Mrk501} we report its observed SED both when only conventional physics is considered and when $\gamma \leftrightarrow a$ oscillations are at work. In order to obtain the SED we take ${\cal E}_{\rm cut}=10 \, \rm TeV$, ${\cal E}_0=1 \, \rm TeV$ and $k=1.8$ in Eq. (\ref{int}).

\item {\it 1ES 0229+200} -- 1ES 0229+200 is a BL Lac observed in the sky at $\rm RA: 02\,h\,32\,m\,48.6\,s$ and $\rm DEC:+20\,d\,17\,m\,17\,s$ at a redshift $z=0.1396$. 
1ES 0229+200 is the prototype of the so-called `extreme HBL' (EHBL)~\citep{btgs2015,cbtgtk2018} which shows a rather hard VHE observed spectrum up to at least 10 TeV. This fact is particularly interesting since the observed data points at such high energies allow to distinguish between different models (conventional physics versus photon-ALP oscillations). Future observations with the CTA that can eventually reach energies up to 100 TeV could give a definitive answer. In Figure~\ref{0229} we plot its observed SED both when only conventional physics is taken into account and in the case in which also $\gamma \leftrightarrow a$ oscillations are present. The SED is obtained by taking in Eq. (\ref{int}) ${\cal E}_{\rm cut}=30 \, \rm TeV$ in the case of conventional physics and ${\cal E}_{\rm cut}=10 \, \rm TeV$ when $\gamma \leftrightarrow a$ oscillations are taken into account, ${\cal E}_0=1 \, \rm TeV$ and $k=1.4$ ($k$ is in agreement with the one derived for the Fermi/LAT spectrum in the recent analysis of~\citealt{cbtgtk2018}). 

\item {\it Extreme BL Lac at $z=0.6$} -- BL Lacs have been observed also at redshift $z \ge 0.6$: we assume the existence of an EHBL at a redshift $z=0.6$. For this blazar we suppose a SED similar to the one of 1ES 0229+200 which is the prototype of EHBLs so that we take ${\cal E}_{\rm cut}=30 \, \rm TeV$, ${\cal E}_0=1 \, \rm TeV$ and $k= 1.4$ in Eq. (\ref{int}). We consider two cases: 1) we imagine that such BL Lac is observed in the sky along the direction of the galactic pole: in Figure~\ref{z06L} we plot its observed SED both when only conventional physics is considered and in the case in which also $\gamma \leftrightarrow a$ oscillations are present; 2) we hypothesize that the same BL Lac is instead observed in the sky along the direction of the galactic plane: in Figure~\ref{z06H} we exhibit the corresponding observed SED according to conventional physics and when $\gamma \leftrightarrow a$ oscillations are taken into account.

\end{itemize}

\section{Results}
Figures~\ref{Mrk501}-\ref{z06H} show our results about the SED of the above-considered BL Lacs. As a general outcome, we get that the $\gamma \leftrightarrow a$ oscillations allow for a harder observed spectra for all sources as compared with the results of conventional physics. In particular, this fact becomes more and more evident as ${\cal E}$ or $z$ (or both) increase.

We infer from our findings that $\gamma \to a$ conversions inside the magnetic field of the BL Lac jet can be very important in order to start the propagation in the extragalactic space with a certain amount of already produced ALPs: its relevance depends both on ${\cal E}$ and on $z$. This point is rather subtle and deserves a clear explanation. Superficially, one might expect $P_{\gamma \to \gamma}^{\rm ALP} ({\cal E}, z)$ to increase with $g_{a \gamma \gamma}$, in line with physical intuition. This is certainly true as long as the EBL does not play an important role, namely for ${\cal E}$ and $z$ low enough. Needless to say, $\gamma \to a$ conversions and $a \to \gamma$ back-conversion in the BL Lac and in the Milky Way help increasing $P_{\gamma \to \gamma}^{\rm ALP} ({\cal E}, z)$, but not that much. Consider next the situation in which both $g_{a \gamma \gamma}$ and $z$ are fairly large but ${\cal E}$ is not, so that photon dispersion on the CMB can be neglected. In such a situation the conversion probability gets enhanced to such an extent that inside a single magnetic domain many $\gamma \to a$ and $a \to \gamma$ conversions take place. But since $z$ is supposed to be rather large the EBL level is high, which causes most of the photons to be absorbed. Such a behaviour is very clearly exhibited in Figures~\ref{z06L} and~\ref{z06H} around ${\cal E} \simeq 3 \, {\rm TeV}$. As the energy increases, photon dispersion on the CMB becomes dominant: now a much smaller number of $\gamma \to a$ and $a \to \gamma$ conversions occurs in the extragalactic space. As a consequence, most of the ALPs produced in the BL Lac survive until they enter the Galaxy, whose strong magnetic field allows them to convert to photons. This fact explains the peak in Figures~\ref{z06L} and~\ref{z06H} around ${\cal E} = (10 - 30) \, {\rm TeV}$. From all the figures we observe that as ${\cal E}$ progressively increases beyond $ 70 \, {\rm TeV}$ the area covered by the various realizations of the photon/ALP propagation process gradually reduces. The reason for this fact is that the EBL absorption is so high at those energies that almost all the photons in each extragalactic magnetic field domain are absorbed and only the ones reconverted from ALPs inside the Galaxy are observed (as previously mentioned). As a result, the parameter space of the model (${\bf B}_{\rm ext}$ orientation angles, domain lengths $L_{\rm dom}$) gets reduced, and this fact decreases the available area that can be covered by the realizations of the propagation process.

In all the figures we report the CTA sensitivity for the South site and 50 h of observation. Since the sensitivity curve is based on conservative criteria~\citep{CTAsens,CTAnew} we expect that the theoretical spectral features (e.g. the peak in Figures~\ref{z06L} and~\ref{z06H} around ${\cal E} \sim 20 \, {\rm TeV}$) which are close to the sensitivity curve should anyhow be detectable by the CTA.

\section{Conclusions}

In this Paper, we have studied the propagation of a photon/ALP beam originating well inside a BL Lac jet and traveling in the jet magnetic field, in the host galaxy magnetic field, in the extragalactic magnetic field, and in the Milky Way magnetic field up to us. We observe from Markarian 501 (see Figure~\ref{Mrk501}) that conventional physics hardly fits the two highest energy points of the SED while the model including $\gamma \leftrightarrow a$ oscillations naturally matches the data. For 1ES 0229+200 (see Figure~\ref{0229}) the model including $\gamma \leftrightarrow a$ oscillations fits well the data, especially concerning the last highest energy data point of the SED. As it is evident from Figures~\ref{z06L} and \ref{z06H} -- as the redshift increases -- at high energies the difference between the results from conventional physics alone, and the model including $\gamma \leftrightarrow a$ oscillations becomes more and more dramatic. This is even more the case when sizable $\gamma \to a$ conversions take place inside a blazar, since then most of the emitted ALPs can become  photons only inside the strong Milky Way magnetic field. In particular, for very distant BL Lacs we predict a peak in the energy spectra at ${\cal E} = (10-30) \, {\rm TeV}$ as it is evident from  Figures~\ref{z06L} and \ref{z06H}. In addition, the energy oscillations in the observed spectrum -- clearly recognizable in the Figures -- are a clear-cut feature of our scenario, which can be observed provided that the detector has enough energy resolution: they arise from the photon dispersion on the CMB.

A competitive scenario capable to reduce the optical depth is the Lorentz invariance violation (LIV) which could predict a somehow similar peak in the BL Lac spectra above $\sim 20 \, \rm TeV$~\citep{LIVstecker,LIVtavecchio}. In any case, the two scenarios can in principle be distinguished since the LIV does not predict any spectral energy oscillatory behaviour.

At this point some remarks are compelling.

\begin{itemize}

\item The jet parameters ($y_{\rm VHE}$, $B_{T, {\cal R}_{\rm VHE}}$) are affected by  uncertainties, and the amount of produced ALPs in this region clearly depends on such quantities. Nevertheless, we have checked that the final spectra qualitatively possess the above-mentioned features regardless of the choice of the jet parameters, provided of course that they are realistic.

\item Even if we consider very low values of the extragalactic magnetic field -- namely  $B_{\rm ext} \ll 10^{- 9} \, \rm G$ -- the considered model predicts the same features even if partially reduced, in particular concerning the amplitude of the energy oscillations. However, the peak in the spectra at ${\cal E} = (10-30) \, {\rm TeV}$ remains unaffected at high redshift.

\item The electromagnetic cascade proposed to mimic photon-ALP oscillation effects in  blazar spectra~\citep{Cascade} can work only for $B_{\rm ext} \lesssim {\cal O}(10^{-15} \, \rm G)$,  which is indeed quite close to the $B_{\rm ext}$ lower limits~\citep{neronov2010,durrerneronov,pshirkov2016}. Still, for $B_{\rm ext} \gtrsim {\cal O}(10^{-15} \, \rm G)$ the charged particles produced in the cascade are deflected by ${\bf B}_{\rm ext}$ and the resulting additional photon flux turns out to be totally irrelevant (for more details, see e.g.~\citealt{noCascade}).

\item For ${\cal E} \gtrsim 100 \, \rm TeV$ the infrared radiation from dust present inside the Milky Way could play a moderate role in absorbing photons~\citep{AbsMW}. But this effect is irrelevant for us and can be safely discarded. The reasons is as follows. The absorption is substantial only inside the Galactic plane and a few degrees above and below it, and so only ALPs converted to photons in the Galactic plane close to the outer border of the Milky Way disk fully undergo such an effect. As a matter of fact, two points should be be stressed. 1) For the line of sight to the blazar outside the galactic plane the considered effect is fully negligible. 2) Even for photons in the photon/ALP beam entering the Milky Way along the Galactic plane the $\gamma \leftrightarrow a$ oscillations reduces photon absorption, thereby making it negligible.

\end{itemize}

Taking into account the above-mentioned remarks, our predictions are of great importance for the new generation of gamma-ray observatories like CTA (Cherenkov Telescope Array)~\citep{cta}, HAWC (High-Altitude Water Cherenkov Observatory)~\citep{hawc}, GAMMA 400 (Gamma-Astronomy Multifunction Modules Apparatus)~\citep{g400}, LHAASO (Large High Altitude Air Shower Observatory)~\citep{lhaaso}, TAIGA-HiSCORE (Tunka Advanced Instrument for Gamma-ray and Cosmic ray Astrophysics-Hundred Square km Cosmic ORigin Explorer)~\citep{desy} and HERD (High Energy cosmic-Radiation Detection)~\citep{herd}, which can test our model and eventually make a first indirect detection of an ALP with properties similar to the ones described in this Paper. We plan to perform dedicated simulations in order to test whether energy oscillations around $500 \, {\rm GeV}-2 \, {\rm TeV}$ in BL Lac spectra are detectable with the CTA, along with the photon excess at $10 - 30 \, {\rm TeV}$.

Still, this is not the end of the story. Because our ALP has mass $m_a = {\cal O}(10^{- 10} \, {\rm eV})$ and assuming that indeed $g_{a \gamma \gamma} \simeq 10^{- 11} \, 
{\rm GeV}^{- 1}$, it can be directly detected in the laboratory within the next few years, thanks to the upgrade of ALPS II at DESY~\citep{alps2}, the planned IAXO~\citep{iaxo,iaxo2} and STAX~\citep{stax} experiments, as well as with other techniques developed by Avignone and collaborators~\citep{avignone1,avignone2,avignone3}. Moreover, if the bulk of the dark matter is made of ALPs they can also be detected by the planned ABRACADABRA experiment~\citep{abracadabra}.

Finally, we plan to consider a much larger number of blazars -- both observed and simulated -- in a more complete and systematic forthcoming publication.

\section*{Acknowledgements}

We thank the referee, Floyd Stecker, for comments that help us to clarify the presentation. G. G. and F. T. acknowledge contribution from the grant INAF CTA--SKA, `Probing particle acceleration and $\gamma$-ray propagation with CTA and its precursors', M. R. acknowledges the financial support by the TAsP grant of INFN and C. E. acknowledges the European Commission for support under the H2020-MSCA-IF-2016 action, Grant No. 751311 `GRAPES Galactic cosmic RAy Propagation: An Extensive Study'.

\label{lastpage}


\begin{thebibliography}{99}

\bibitem[\protect\citeauthoryear{Acharyya et al.}{2019}]{CTAnew} Acharyya A. {\it et al.}, Astroparticle Physics {\bf 111}, 35-53 (2019).

\bibitem[\protect\citeauthoryear{Aharonian et al.}{2001}]{hegra} Aharonian F. {\it et al.}, Astrophys. J. {\bf 546}, 898 (2001). 

\bibitem[\protect\citeauthoryear{Aharonian et al.}{2007}]{1eshess} Aharonian F. {\it et al.} [H.E.S.S. Collaboration], Astron. Astrophys. {\bf 475}, L9 (2007).

\bibitem[\protect\citeauthoryear{Ajello et al.}{2016}]{fermi2016} Ajello M. {\it et al}., [Fermi-LAT collaboration], Phys. Rev. Lett. {\bf 116}, 161101 (2016).

\bibitem[\protect\citeauthoryear{Anastassopoulos et al.}{2017}]{cast} Anastassopoulos V. {\it et al.} [CAST Collaboration], Nature Physics {\bf 13}, 584 (2017).

\bibitem[\protect\citeauthoryear{Anselm}{1985}]{anselm} Anselm A. A., Yad. Fiz. {\bf 42}, 1480 (1985).

\bibitem[\protect\citeauthoryear{Armengaud et al.}{2019}]{iaxo2} Armengaud E. {\it et al.}, arXiv:1904.09155.

\bibitem[\protect\citeauthoryear{Avignone}{2009}]{avignone1} Avignone III F. T., Phys. Rev. D {\bf 79}, 035015 (2009).

\bibitem[\protect\citeauthoryear{Avignone, Crewick \& Nussinov}{2009}]{avignone2} Avignone III F. T., Crewick R. J. and Nussinov S., Phys. Lett. B {\bf 681}, 122 (2009).

\bibitem[\protect\citeauthoryear{Avignone, Crewick \& Nussinov}{2011}]{avignone3} Avignone III F. T., Crewick R. J. and Nussinov S., Astropart. Phys. {\bf 34}, 640 (2011). 

\bibitem[\protect\citeauthoryear{Ayala et al.}{2014}]{straniero} Ayala A. {\it et al.}, Phys. Rev. Lett. {\bf 113}, 191302 (2014).

\bibitem[\protect\citeauthoryear{B\"ahre et al.}{2013}]{alps2} B\"ahre R. {\it et al.}, J. of Instrum. {\bf 8}, T09001 (2013). 

\bibitem[\protect\citeauthoryear{Begelman, Blandford \& Rees}{1984}]{bbr1984} Begelman M. C., Blandford R. D., Rees M. J., Rev. Mod. Phys. {\bf 56}, 255 (1984).

\bibitem[\protect\citeauthoryear{Bernl\"ohr et al.}{2013}]{CTAsens} Bernl\"ohr K. {\it et al.}, APh, {\bf 43}, 171 (2013).

\bibitem[\protect\citeauthoryear{Bonnoli, Tavecchio, Ghisellini \& Sbarrato}{2015}]{btgs2015} Bonnoli G., Tavecchio F., Ghisellini G. and Sbarrato T., Mon. Not. R. Astron. Soc. {\bf 451}, 611 (2015).

\bibitem[\protect\citeauthoryear{Breit \& Wheeler}{1934}]{breitwheeler} Breit G. and Wheeler J. A., Phys. Rev. {\bf 46}, 1087 (1934).

\bibitem[\protect\citeauthoryear{Capparelli et al.}{2016}]{stax} Capparelli L. M. {\it et al.}, Phys. Dark Univ. {\bf 12}, 37 (2016).

\bibitem[\protect\citeauthoryear{Costamante et al.}{2018}]{cbtgtk2018} Costamante L., Bonnoli G., Tavecchio F., Ghisellini G., Tagliaferri G. and Khangulyan D., Mon. Not. R. Astron. Soc. {\bf 477}, 4257 (2018).

\bibitem[\protect\citeauthoryear{CTA}{website}]{cta} CTA, https://www.cta-observatory.org/

\bibitem[\protect\citeauthoryear{De Angelis, Galanti \& Roncadelli}{2011}]{dgrx} De Angelis A., Galanti G. and Roncadelli M., Phys. Rev. D {\bf 84}, 105030 (2011); (E) ibid. {\bf 84} 105030 (2013).

\bibitem[\protect\citeauthoryear{De Angelis, Galanti \& Roncadelli}{2013}]{dgr2013} De Angelis A., Galanti G. and Roncadelli M., Mon. Not. R. Astron. Soc. {\bf 432}, 3245 (2013).

\bibitem[\protect\citeauthoryear{De Angelis, Roncadelli \& Mansutti}{2007}]{drm2007} De Angelis A., Roncadelli M. and Mansutti O., Phys. Rev. D {\bf 76}, 121301 (2007).

\bibitem[\protect\citeauthoryear{Dobrynina, Kartavtsev \& Raffelt}{2015}]{raffelt2015} Dobrynina A., Kartavtsev A. and Raffelt G., Phys. Rev. D {\bf 91} (2015) 083003; {\bf 91}, 109902 (E) (2015).

\bibitem[\protect\citeauthoryear{Durrer \& Neronov}{2013}]{durrerneronov} Durrer R. and Neronov A., Astron. Astrophys. Rev. {\bf 21}, 62 (2013).

\bibitem[\protect\citeauthoryear{Dwek}{2013}]{dwek}  Dwek E. and Krennrich F., Astropart. Phys. {\bf 43}, 112 (2013).

\bibitem[\protect\citeauthoryear{Dzhatdoev, Khalikov, Kircheva \& Lyukshin}{2017}]{Cascade} Dzhatdoev T. A., Khalikov E. V., Kircheva A. P. and Lyukshin A. A., Astron. Astrophys. {\bf 603}, A59 (2017).

\bibitem[\protect\citeauthoryear{Fazio \& Stecker}{1970}]{stecker1970} Fazio G. G. and Stecker F. W., Nature {\bf 226}, 135 (1970).

\bibitem[\protect\citeauthoryear{Franceschini \& Rodighiero}{2017}]{franceschinirodighiero} Franceschini A. and Rodighiero G., Astron. Astrophys. {\bf 603}, 34 (2017).   

\bibitem[\protect\citeauthoryear{Furlanetto \& Loeb}{2001}]{furlanettoloeb} Furlanetto S. and Loeb A., Astrophys. J. {\bf 556}, 619 (2001)

\bibitem[\protect\citeauthoryear{Galanti \& Roncadelli}{2018a}]{grSM} Galanti G. and Roncadelli M., Phys. Rev. D {\bf 98}, 043018 (2018).

\bibitem[\protect\citeauthoryear{Galanti \& Roncadelli}{2018b}]{grExt} Galanti G. and Roncadelli M., JHEAp {\bf 20}, 1 (2018).

\bibitem[\protect\citeauthoryear{GAMMA 400}{website}]{g400} GAMMA 400, gamma400.lebedev.ru/gamma400e.html.

\bibitem[\protect\citeauthoryear{Ghisellini \& Tavecchio}{2009}]{ghisellini2009} Ghisellini G., Tavecchio F., Mon. Not. R. Astron. Soc. {\bf 397}, 985  (2009).

\bibitem[\protect\citeauthoryear{Gould \& Schr\'eder}{1967}]{gs1967} Gould R. J. and Schr\'eder G. P., Phys. Rev. {\bf 155}, 1408 (1967).

\bibitem[\protect\citeauthoryear{Grasso \& Rubinstein}{2001}]{grassorubinstein} Grasso D. and Rubinstein H. R., Phys. Rep. {\bf 348}, 163 (2001).

\bibitem[\protect\citeauthoryear{HAWC}{website}]{hawc} HAWC, www.hawc-observatory.org/.

\bibitem[\protect\citeauthoryear{Heisenberg \& Euler}{1936}]{HEW1} Heisenberg W., Euler H., Z. Phys. {\bf 98}, 714-732 (1936).

\bibitem[\protect\citeauthoryear{Heitler}{1960}]{heitler} Heitler W., {\it The Quantum Theory Of Radiation} (Oxford University Press, Oxford, 1960).

\bibitem[\protect\citeauthoryear{HESS}{website}]{hess} HESS, https://www.mpi-hd.mpg.de/hfm/HESS

\bibitem[\protect\citeauthoryear{Horns et al.}{2012}]{hmmmmr2012} Horns D., Maccione L., Meyer M., Mirizzi A., Montanino D., and Roncadelli M., Phys. Rev. D {\bf 86}, 075024 (2012).

\bibitem[\protect\citeauthoryear{Hoyle}{1969}]{hoyle} Hoyle F., Nature {\bf 223}, 936 (1969).

\bibitem[\protect\citeauthoryear{Huang et al.}{2016}]{herd} Huang X. {\it et al.}, Astropart. Phys. {\bf 78}, 35 (2016).

\bibitem[\protect\citeauthoryear{Irastorza et al.}{2011}]{iaxo} Irastorza I. G. {\it et al.} [IAXO Collaboration], JCAP {\bf 06}, 013 (2011).

\bibitem[\protect\citeauthoryear{Jansson \& Farrar}{2012a}]{jansonfarrar1} Jansson R. and Farrar G. R., Astrophys. J. {\bf 757}, 14 (2012). 

\bibitem[\protect\citeauthoryear{Jansson \& Farrar}{2012b}]{jansonfarrar2} Jansson R. and Farrar G. R.,  Astrophys. J. {\bf 761}, L11 (2012).

\bibitem[\protect\citeauthoryear{Jaeckel \& Ringwald}{2010}]{alp1} Jaeckel J. and Ringwald A., Ann. Rev. Nucl. Part. Sci. {\bf 60}, 405 (2010).

\bibitem[\protect\citeauthoryear{Kahn, Safdi \& Thaler}{2016}]{abracadabra} Kahn Y., Safdi B. R. and Thaler J., Phys. Rev. Lett. 117, 141801 (2016).

\bibitem[\protect\citeauthoryear{Kartavtsev, Raffelt \& Vogel}{2017}]{raffeltvogel} Kartavtsev A., Raffelt G. and Vogel H., JCAP {\bf 01}, 024 (2017).

\bibitem[\protect\citeauthoryear{Kohri \& Kodama}{2017}]{kk2017} Kohri K. and Kodama H., Phys. Rev. D {\bf 96}, 051701 (2017).

\bibitem[\protect\citeauthoryear{Kronberg}{1994}]{kronberg1994} Kronberg P. P., Rept. Prog. Phys. {\bf 57}, 325 (1994).

\bibitem[\protect\citeauthoryear{Kronberg, Lesch \& Hopp}{1999}]{kronberg1999} Kronberg P. P., Lesch H. and Hopp U., Astrophys. J. {\bf 511}, 56 (1999).

\bibitem[\protect\citeauthoryear{LHAASO}{website}]{lhaaso} LHAASO, http://english.ihep.cas.cn/ic/ip/LHAASO/.

\bibitem[\protect\citeauthoryear{MAGIC}{website}]{magic} MAGIC, https://magic.mpp.mpg.de/

\bibitem[\protect\citeauthoryear{Masaki, Aoki \& Soda}{2019}]{jap} Masaki E., Aoki A. and Soda J., arxiv:1702.08843.

\bibitem[\protect\citeauthoryear{Matsuura et al.}{2017}]{ciber} Matsuura S. {\it et al.}, Astrophys. J. {\bf 839}, 7 (2017). 

\bibitem[\protect\citeauthoryear{Moss \& Shukurov}{1996}]{moss1996} Moss D. and Shukurov A., Mon. Not. R. Astron. Soc. {\bf 279}, 229 (1996).

\bibitem[\protect\citeauthoryear{Neronov \& Vovk}{2010}]{neronov2010} Neronov A. and Vovk I., Science {\bf 328}, 73 (2010).

\bibitem[\protect\citeauthoryear{Payez et al.}{2015}]{payez2015} Payez A. {\it et al.}, JCAP {\bf 02}, 006 (2015).

\bibitem[\protect\citeauthoryear{Pshirkov, Tinyakov, Kronberg \& Newton-McGee}{2011}]{pshirkovMF2011} Pshirkov M. S., Tinyakov P. G., Kronberg P. P. and Newton-McGee K. J., Astrophys. J. 738, 192 (2011).

\bibitem[\protect\citeauthoryear{Pshirkov, Tinyakov \& Urban}{2016}]{pshirkov2016} Pshirkov M. S., Tinyakov P. G. and Urban F. R., Phys. Rev. Lett. {\bf 116}, 191302 (2016).

\bibitem[\protect\citeauthoryear{Pudritz, Hardcastle \& Gabuzda}{2012}]{pudritz2011} Pudritz R. E., Hardcastle M. J. and Gabuzda D. C., Space Sci. Rev. {\bf 169}, 27 (2012).

\bibitem[\protect\citeauthoryear{Raffelt \& Stodolsky}{1988}]{rs1988} Raffelt G. and Stodolsky L., Phys. Rev. D {\bf 37}, 1237 (1988).

\bibitem[\protect\citeauthoryear{Rees \& Setti}{1968}]{reessetti} Rees M. J. and Setti G., Nature {\bf 219}, 127 (1968).

\bibitem[\protect\citeauthoryear{Ringwald}{2012}]{alp2} Ringwald A., Phys. Dark Univ. {\bf 1}, 116 (2012).

\bibitem[\protect\citeauthoryear{S\'anchez-Conde et al.}{2009}]{prada1} S\'anchez-Conde M. A. {\it et al}., Phys. Rev. D {\bf 79}, 123511 (2009).

\bibitem[\protect\citeauthoryear{Schwinger}{1951}]{HEW3} Schwinger J., Phys. Rev. {\bf 82}, 664-679 (1951).

\bibitem[\protect\citeauthoryear{Sikivie}{1984}]{sikivie} Sikivie P., Phys. Rev. Lett. {\bf 51}, 1415 (1983); (E) ibid. {\bf 52}, 695 (1984).

\bibitem[\protect\citeauthoryear{Simet, Hooper \& Serpico}{2008}]{shs2008} Simet M., Hooper D. and Serpico P. D., Phys. Rev. D {\bf 77}, 063001 (2008).

\bibitem[\protect\citeauthoryear{Stecker \& Glashow}{2001}]{LIVstecker} Stecker, F. W., and Glashow, S. L., Astroparticle Physics, {\bf 16}, 97 (2001).

\bibitem[\protect\citeauthoryear{TAIGA-HiSCORE}{website}]{desy} TAIGA-HiSCORE, www.desy.de/groups/astroparticle/score/en/.

\bibitem[\protect\citeauthoryear{Tavecchio \& Bonnoli}{2016}]{LIVtavecchio} Tavecchio F. and Bonnoli G., \aap, {\bf 585}, A25 (2016).

\bibitem[\protect\citeauthoryear{Tavecchio et al.}{2010a}]{tavecchio2010} Tavecchio F., Ghisellini G, Ghirlanda G., {\it et al.}, Mon. Not. R. Astron. Soc. {\bf 401}, 1570  (2010).

\bibitem[\protect\citeauthoryear{Tavecchio et al.}{2010b}]{noCascade} Tavecchio F., Ghisellini G., Foschini L., {\it et al.}, Mon. Not. R. Astron. Soc. {\bf 406}, L70 (2010).

\bibitem[\protect\citeauthoryear{Tavecchio, Roncadelli \& Galanti}{2015}]{trg} Tavecchio F., Roncadelli M. and Galanti G., Phys. Lett. B {\bf 744}, 375 (2015).

\bibitem[\protect\citeauthoryear{Tavecchio, Roncadelli, Galanti \& Bonnoli}{2012}]{trgb2012} Tavecchio F., Roncadelli M., Galanti G. and Bonnoli G., Phys. Rev. D {\bf 86}, 085036 (2012).

\bibitem[\protect\citeauthoryear{Unger \& Farrar}{2019}]{uf2017} Unger M. and Farrar G. R., arXiv:1707.02339. 

\bibitem[\protect\citeauthoryear{VERITAS}{website}]{veritas} VERITAS, http://veritas.sao.arizona.edu/

\bibitem[\protect\citeauthoryear{Vernetto \& Lipari}{2016}]{AbsMW} Vernetto S. and Lipari P., Phys. Rev. D {\bf 94}, 063009 (2016).

\bibitem[\protect\citeauthoryear{Vovk, Taylor, Semikoz \& Neronov}{2012}]{1esfermi} Vovk I., Taylor A. M., Semikoz D. and Neronov A., Astrophys. J. {\bf 747}, L14 (2012).

\bibitem[\protect\citeauthoryear{Wang \& Lai}{2016}]{wanglai} Wang C. and Lai D., JCAP {\bf 16}, 006 (2016).

\bibitem[\protect\citeauthoryear{Weisskopf}{1936}]{HEW2} Weisskopf V. S., K. Dan. Vidensk. Selsk. Mat. Fys. Medd. {\bf 14}, 6 (1936).

\bibitem[\protect\citeauthoryear{Yao, Manchester \& Wang}{2017}]{ymw2017} Yao J. M., Manchester R. N. and Wang N.,  Astrophys. J. {\bf 835}, 29 (2017).

\end{thebibliography}
\end{document}